\documentclass[preprint,aps]{revtex4}

\usepackage{graphicx}
\usepackage{dcolumn}
\usepackage{bm}

\newcommand{\be}{\begin{equation}}
\newcommand{\ee}{\end{equation}}
\newcommand{\bey}{\begin{eqnarray}}
\newcommand{\eey}{\end{eqnarray}}
\newcommand{\bw}{\begin{widetext}}
\newcommand{\ew}{\end{widetext}}

\makeatletter
    
    \newcommand{\Rmnum}[1]{\expandafter\@slowromancap\romannumeral #1@}
\makeatother

\begin{document}

\title{Time Reversal Symmetry Breaking Holographic
Superconductor in Constant External Magnetic Field}
\author{Huabi Zeng,
 Zheyong Fan,
 and Zhongzhou Ren}
\address{Department of Physics, Nanjing University,
Nanjing 210093, China}

\begin{abstract}
It is known that a classical $SU(2)$ Einstein-Yang-Mills theory in
3+1 dimensional anti-de Sitter spacetime can provide a holographic
dual to a 2+1 dimensional time reversal symmetry breaking
superconductor with a pseudogap. We study the properties of this
holographic superconductor in the presence of an applied constant
external magnetic field, neglecting backreaction on the geometry.
The superconductor is immersed into a constant external magnetic
field by adding a radially (the extra dimension) dependent magnetic
field to the black hole. As for real superconductors, there is a
critical magnetic field above which no superconductivity can appear.
The continuity of the first derivative of the free energy difference
between the superconducting phase and the normal phase at the
critical temperature suggests that the superconducting phase
transition with applied magnetic field is of second order.
\end{abstract}
 \maketitle

\section{ introduction}

As a powerful tool to understand strong coupled gauge theories,
AdS/CFT correspondence \cite{maldacena,gubser1,witten,aharony} has
been applied in condensed matter physics recently. There are many
attempts to use this gauge/gravity correspondence to describe
certain condensed matter phenomenons such as the (classical) Hall
effect \cite{hartnoll1}, the Nernst effect
\cite{hartnoll2,hartnoll3,hartnoll4}, the  quantum Hall effect
\cite{keski,davis,fujita}, superconductivity
\cite{gubser2,hartnoll5,gubser3,nakano,albash,gubser4,roberts,
      maeda,hartnoll6,ammon1,basu,herzog1,ammon2,sonner}
and superfluid \cite{herzog2}.

Superconductivity is a common phenomenon occurring in certain
materials, characterized by exact zero electrical resistance and the
exclusion of the interior magnetic field (the Meissner effect). The
essence of superconductivity is the spontaneous breaking of a local
$U(1)$ gauge symmetry due to a charged condensate, which is formed
by Cooper pairs of electrons in the BCS theory. Models of black holes coupled to matter fields also exhibit spontaneous symmetry breaking solutions. Gubser \cite{gubser2} has presented an argument that by coupling the Abelian Higgs model to gravity with a negative cosmological constant, one gets solutions which spontaneously break the Abelian gauge symmetry via a charged complex scalar condensate near the horizon of the black hole. An Einstein-Yang-Mills (EYM) model with fewer
parameters whose Lagrangian is mostly determined by symmetry
principles is constructed later by Gubser \cite{gubser3} and is
shown to have spontaneous symmetry breaking solutions due to a
condensate of non-Abelian gauge fields in the theory. Hartnoll {\it
et al} \cite{hartnoll5} explored further the connections of
superconductors and black holes and built a holographic (in the
sense of AdS/CFT duality) superconductor which exhibits the basic
features of a superconductor such as the existence of a critical
temperature below which a charged condensate forms.

Features that are not those of BCS theory are also captured by
this holographic description of superconductivity. Gubser and Pufu
\cite{gubser4} studied a model of superconducting black holes with
$p$-wave gap solutions whose order parameter is a vector. Roberts
and Hartnoll \cite{roberts} found two major nonconventional
features for the holographic superconductor whose dual theory
involves a classical $SU(2)$ EYM theory, one of which is a
pseudogap at zero temperature, and the other is the spontaneous
breaking of time reversal invariance.

While the spontaneous breaking of time reversal invariance is
exhibited by a nonvanishing Hall conductivity  in the absence of
an external magnetic field,  our purpose in this paper is to study
the behavior of this time reversal symmetry breaking holographic
superconductor in the presence of a constant external magnetic
field. This external magnetic field is implemented on the
superconductor  by considering a magnetically charged
Reissner-Nordstr\"{o}m black hole rather than a Schwarzchild black
hole in anti-de Sitter (AdS) spacetime in the gravity sector. Our
numerical investigations show that the critical temperature
decreases with the increasing of the magnetic field and there is a
critical magnetic field above which no phase transition occurs.
The phase transition at the critical temperature is a second order
one, as suggested by the continuity of the first derivative of the
free energy difference between the superconducting phase and the
normal phase.

The organization of this paper is as follows. In Sec. \Rmnum{2}, we first review the EYM theory which is dual to the
time reversal symmetry breaking superconductor and then discuss
the effect of adding an external magnetic field to it. We obtain in
this section analytically the negative effective mass of the
condensate, which is a sign of spontaneous symmetry breaking
solutions to the model. Section \Rmnum{3} is devoted to the numerical
studies of this model. Section \Rmnum{4} concludes and gives some
further discussions.

\section{ Model of holographic superconductor
in external magnetic field}

The effective physics of many unconventional superconductors such
as cuprates is in 2+1 dimensional spacetime. We will focus in this
paper on a 2+1 dimensional superconductor system which is a gauge
field theory whose holographic dual is a 3+1 dimensional theory
with gravity in AdS spacetime
($\textmd{AdS}_4/\textmd{CFT}_3$).

The starting point of studying the holographic superconductor at finite
temperature $T$ is choosing a black hole solution with a negative
cosmological constant so that the Hawking temperature of the black
hole is $T$. The full action of the EYM theory in 3+1
dimensional spacetime considered in \cite{roberts} consists of
two sectors, the gravity sector and the matter sector,
\begin{equation}
S_{\rm EYM}=
\int\sqrt{-g}d^4x\left[\frac{1}{2\kappa_4^2}\left(R+\frac{6}{L^2}\right)
-\frac{1}{2g_{\rm YM}^2}\textmd{Tr}(F_{\mu\nu}F^{\mu\nu})\right],
\end{equation}
where $g_{YM}$ is the gauge coupling constant
and $F_{\mu\nu}=T^aF^a_{\mu\nu}=\partial_\mu A_\nu-\partial_\nu
A_\mu-i[A_\mu,A_\nu]$ is the  field strength of the gauge field
$A=A_\mu dx^\mu=T^aA^a_\mu dx^\mu$. For the $SU(2)$ gauge symmetry,
$[T^a,T^b]=i\epsilon^{abc}T^c$ and
$\textmd{Tr}(T^aT^b)=\delta^{ab}/2$, where $\epsilon^{abc}$ is the
totally antisymmetric tensor with $\epsilon^{123}=1$. The Yang-Mills
Lagrangian becomes
$\textmd{Tr}(F_{\mu\nu}F^{\mu\nu})=F^a_{\mu\nu}F^{a\mu\nu}/2$ with
the field strength components $F^a_{\mu\nu}=\partial_\mu
A^a_\nu-\partial_\nu A^a_\mu+\epsilon^{abc}A^b_\mu A^c_\nu$.

Working in the probe limit in which the matter fields do not back
react on the metric as in \cite{gubser4,roberts} and
taking the planar Schwarzchild-AdS  ansatz, the  black hole metric
reads (We use mostly plus signature for the metric.)
\begin{equation}
ds^2=-f(r)dt^2+\frac{dr^2}{f(r)}+r^2(dx_1^2+dx_2^2),
\end{equation}
where the metric function $f(r)$ is
\begin{equation}
f(r)=\frac{r^2}{L^2}-\frac{M}{r}.
\end{equation}
$L$ and $M$ are the radius of the AdS spacetime and the mass of
the black hole, respectively. They determine the Hawking
temperature of the black hole,
\begin{equation}
T=\frac{3M^{1/3}}{4\pi L^{4/3}},
\end{equation}
which is also the temperature of the dual gauge theory living on
the boundary of the AdS spacetime.

The equations of motion for the gauge fields can be obtained by
using the Euler-Lagrange equations to be
\begin{equation}
\frac{1}{\sqrt{-g}}\partial_{\mu}\left(\sqrt{-g}F^{a\mu\nu}\right)
+\epsilon^{abc}A^{b}_{u}F^{c\mu\nu}=0.
\end{equation}
Considering the ansatz \cite{gubser3,roberts}
\begin{equation}
A=\phi(r)T^3dt+w(r)(T^1dx^1+T^2dx^2)
\end{equation}
for the gauge fields, the Yang-Mills Lagrangian density reads
[Note that $\sqrt{-g}=r^2$, as can be seen from (2).]
\begin{equation}
\begin{array}{ll}
\mathcal{L}_{\rm YM}=
-\frac{\sqrt{-g}}{2g_{\rm YM}^2}\textmd{Tr}\left(F_{\mu\nu}F^{\mu\nu}\right)
=-\frac{1}{4g_{\rm YM}^2}
\left(-\frac{2}{f}\phi^2w^2+2fw'^{2}-r^2\phi'^{2}+\frac{w^4}{r^2}\right).
\end{array}
\end{equation}
From this Lagrangian density, we can derive the following
equations of motion,
\begin{equation}
\phi''+\frac{2}{r}\phi'-\frac{2w^2}{r^2f}\phi=0,
\end{equation}
and
\begin{equation}
w''+\frac{f'}{f}w'+\frac{\phi^2}{f^2}w-\frac{1}{r^2f}w^3=0,
\end{equation}
which can also be deduced from (5) and (6).

The $U(1)$ subgroup of $SU(2)$ generated by $T^3$ is identified to
be the electromagnetic gauge group \cite{gubser3} and $\phi$ is
the electrostatic potential, which must vanish at the horizon
for the gauge field $A$ to be well defined, but need not vanish at infinity.
Thus the black hole can carry charge through the condensate $w$, which
spontaneously breaks the $U(1)$ gauge symmetry. From the Yang-Mills Lagrangian density (7),
we can see that the effective mass (along the radial direction $r$) for $w$ is
\begin{equation}
m_{eff}^2=-\phi^2/f<0,
\end{equation}
which is characteristic to spontaneous symmetry breaking theories.

In this paper, we are interested in the properties of this kind of
holographic superconductor with an applied external magnetic field.
To achieve this, we consider a magnetically charged
Reissner-Nordstr\"{o}m black hole rather than a Schwarzchild-AdS
black hole in anti-de Sitter space-time (RNAdS) in the gravity
sector \cite{romans,nakano,albash}. Now the full EYM action becomes
\begin{equation}
S'_{\rm EYM}=
\int\sqrt{-g}d^4x\left[\frac{1}{2\kappa_4^2}\left(R+\frac{6}{L^2}\right)-\frac{1}{4}\mathcal
{F}_{\mu\nu} \mathcal{F}^{\mu\nu}
-\frac{1}{2g_{\rm YM}^2}\textmd{Tr}(F_{\mu\nu}F^{\mu\nu})\right],
\end{equation}
where the first two terms belong to the gravity sector. By
setting the nonvanishing components of the field strength of the Maxwell field to be
$\mathcal {F}_{xy}=r^2H=-\mathcal {F}_{yx}$ such that $-\mathcal
{F}_{\mu\nu} \mathcal{F}^{\mu\nu}/4=-H^2/2$, we immerse the
superconductor at the AdS boundary into a perpendicular
constant external magnetic field.

We still work in the probe limit \cite{gubser4,roberts} where the metric is only determined
by the gravity sector. In this limit, the background geometry is given by the magnetically charged
RNAdS black hole with a metric taking the same form as in (2), but with the metric function (3) being replaced by
\begin{equation}
f(r)=\frac{r^2}{L^2}-\frac{M}{r}+\frac{H^2}{r^2}.
\end{equation}
The Hawking temperature of the RNAdS black hole
is determined by
\begin{equation}
T=\frac{f'(r_+)}{4\pi},
\end{equation}
where $r_+$ is the radius of the outer horizon of the
RNAdS black hole. We notice that for the vanishing
magnetic field $H=0$, the metric function $f(r)$ in (12) and the
Hawking temperature $T$ in (13) reduce to the previous ones in (3)
and (4), respectively.

In the probe limit, by adding an external magnetic field, the black hole background considered is changed, and the equations of motion for the non-Abelian gauge fields have the same forms as in (8) and (9) with the metric function $f(r)$ replaced by (12). Our task in the
next section is to find solutions to the equations of motion (8) and
(9) of the fields $\phi$ and $w$ with appropriate boundary
conditions to see how the applied magnetic field affects the dual superconductor.

\section{ Solutions to the model}
There are two solutions to the equations of motion of the fields
$\phi$ and $w$, one without hair and the other with hair,
corresponding to the normal and the superconductor states of the
system, respectively.

The no hair solution is easy to find by setting $w=0$, which gives
$\phi=\mu-\rho/r$, where $\mu$ and $\rho$ are the chemical potential
and the charge density of the field theory, respectively.

The hairy solution can only be found by numerically solving the
second order nonlinear equations (8) and (9) with the boundary
condition $\phi(r_+)=0$ for $\phi(r)$ at the outer horizon and the
following asymptotic behaviors of the fields at the AdS conformal
boundary $r\rightarrow\infty$ \cite{roberts},
\begin{equation}
\phi=\mu-\frac{\rho}{r}+\cdots,
\end{equation}
\begin{equation}
w=w_0+\frac{w_1}{r}+\cdots = \frac{\langle
J\rangle}{\sqrt{2}r}+\cdots,
\end{equation}
where $\langle J\rangle$ is the condensate  of the charged
operator dual to the field $w$ and is the order parameter for the
superconductivity phase. The constant term $w_0$ vanishes, since
there is no source term in the field theory action for the operator
$\langle J \rangle$ \cite{gubser4,roberts}.

Properties of the dual field theory can be read off from the above
asymptotic behaviors via the AdS/CFT correspondence. The strategy
to find the above hairy black hole solution is to expand the
fields near the outer horizon and numerically integrate them out
from the horizon to infinity. To facilitate the numerical
calculations, we set the AdS radius to be $L=1$, which can be
achieved by using the rescaling symmetry of the model. In the
absence of a magnetic field, there is only one horizon $r_0$, which
can be set to be 1 by the conformal properties, and one can
develop power series solutions for $\phi$ and $w$ near the horizon
$z=1$ where $z$ is the inverse of $r$. This change of variable
from $r$ to $z$ brings much convenience to numerical studies. However, with the
introduction of the magnetic field, the outer horizon sits at
$z=\frac{1}{r_+}>1$, making the series expansions of the fields
near the outer horizon much more complicated and disabling
Mathematica for doing the subsequent numerical calculations. We
circumvent this difficulty by using another change of variable
from $z$ to $y=r_+z$ such that power series solutions for $\phi$
and $w$ can be developed near the horizon $y=1$.

\begin{figure}[!htbp] \centering
\includegraphics[width=3.0in]{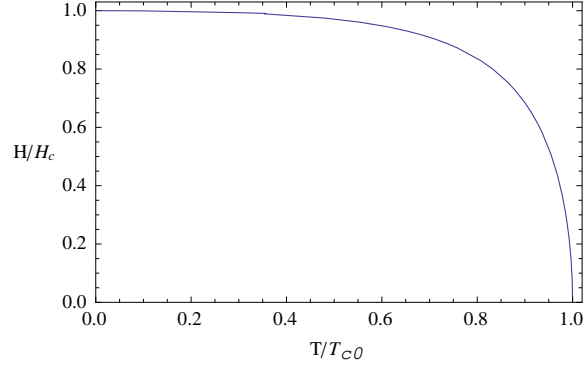}
\hspace{2.2cm}\caption{Dependence of the critical temperature
$T_c$ (along the line in the diagram) on the applied magnetic
field $H$. $T_{c0}$ and $H_c$ are the critical temperature without
the applied magnetic field ($H=0$) and the critical magnetic field at
zero temperature, respectively.}
\end{figure}

\begin{figure}[!htbp] \centering
\includegraphics[width=3.0in]{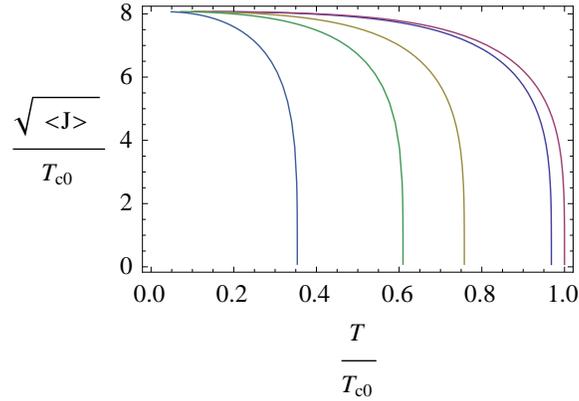}
\hspace{2.2cm}\caption{The order parameters $\sqrt{\langle J
\rangle}$ as functions of the temperature for different magnitudes of
the applied magnetic field $H$. $T_{c0}$ is the critical temperature
without the applied magnetic field. The lines from right to left
correspond to the cases $H=0$, 0.3, 0.6, 0.65, and 0.68,
respectively.}
\end{figure}

We have studied systematically the numerical solutions with
different magnitudes of the magnetic field $H$. We found that
there is a hair solution for any $H$ in the range $0\leq H<
H_{\textmd{max}}$, where $H_{\textmd{max}}=0.687365$ is the
solution to the equation $27-256H^6=0$ and is the maximal value
the magnetic field can take in order for the black hole to be
censored by at least one horizon. It turns out that the critical
temperature $T_c$ decreases with the increasing of the magnetic
field $H$. See Figs. 1 and 2 for  illustrations. The
decreasing of the $T_c$ is very slow when  $H$ is small and gets
more and more rapid when $H$ becomes lager and larger. Finally,
$T_c$ vanishes when $H$ reaches the critical value
$H_c=H_{\rm max}$.

Important information about the phase transition can be extracted
from the behavior of the free energy, as in the studies on
holographic superfluidity by Herzog \it et al \rm \cite{herzog2}.
The free energy of the field theory is determined by the value of
the Yang-Mills action (ignoring the backreaction of the gauge
fields on the metric)
\begin{equation}
S_{\rm YM}= \int d^4x\mathcal{L}_{\rm YM}
\end{equation}
evaluated on shell up to boundary counterterms,
$F=-TS_{\rm os}+\cdots$, where the ellipsis denotes boundary
terms that we should introduce to regulate the action if needed.
The on-shell Yang-Mills action $S_{\rm os}$ is determined by
plugging the equations of motion (8) and (9) into the explicit
form of the Yang-Mills Lagrangian (7) (omitting the irrelevant
factor $1/4g_{\rm YM}^2$),
\begin{equation}
\begin{array}{ll}
S_{\rm os}=\int{d^3x}(-\phi\phi'+2z^2fww')|_{z=\epsilon}
-\int{d^3x}\int^{z_h}_{\epsilon}dz
\left(-w^4+\frac{2\phi^2w^2}{fz^2}\right),
\end{array}
\end{equation}
where the metric function $f$ in (17) should be understood as a
function of the inverse radial (or Fefferman-Graham) coordinate $z=1/r$, $f(z)=1/L^2z^2-zM+H^2z^2$ and $\epsilon = 0^+$ and $z_h$ are the boundary of the AdS spacetime and the outer horizon of the RNAdS black hole, respectively.

To regulate $S_{\rm os}$, it is important to choose an
ensemble. By keeping $\mu$ fixed, we are working in the grand
canonical ensemble without an additional boundary term
\cite{herzog2}. Near the boundary $z=\epsilon$, the fields $\phi$
and $w$ are determined by (14) and (15) and the two terms $-\phi
\phi'$ and  $2z^2fww'$ in (17) give $\mu \rho$ and $2w_0w_1$,
respectively. We can see that the on-shell action
$S_{\rm os}$ is not divergent and no counterterms are
needed. Since $w_0$ is fixed to be zero, for a spatially
homogenous system, the free energy density of the field theory
takes the following form,
\begin{equation}
F/V= -\mu\rho
+\int^{z_h}_{\epsilon}dz\left(-w^4+\frac{2\phi^2w^2}{fz^2}\right),
\end{equation}
where $V\equiv\int{d^3x}$.

\begin{figure}[!htbp] \centering
\includegraphics[width=3.0in]{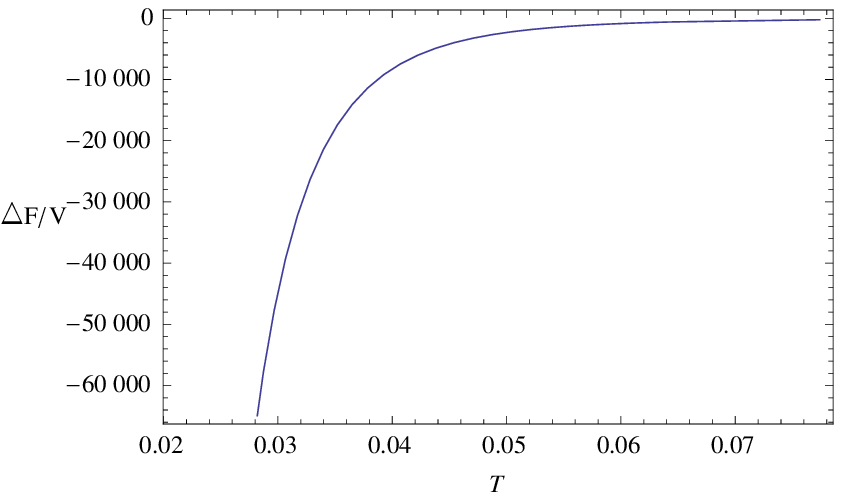}
\includegraphics[width=3.0in]{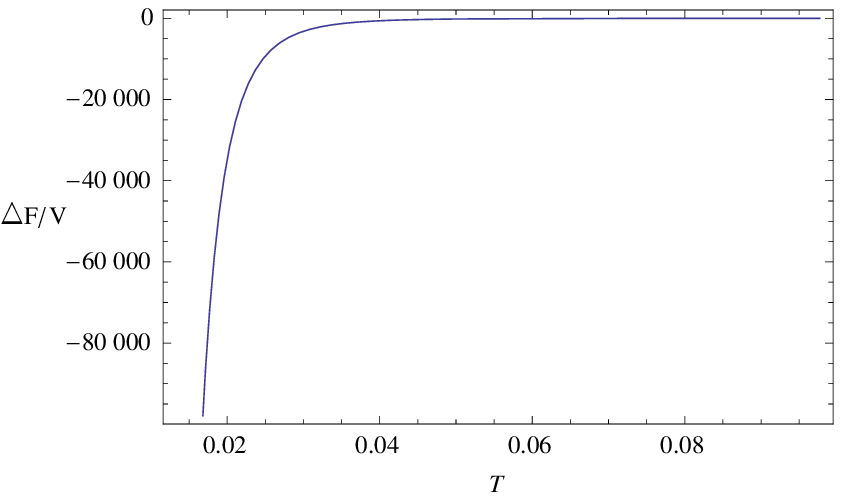}
\hspace{2.2cm}\caption{The difference in free energy density
$\Delta F/V$ between the superconducting phase and the normal
phase as a function of the temperature $T$ for $H=0.2$ (left
diagram) and $H=0.6$ (right diagram).}
\end{figure}

Figure 3 shows the difference of the free energy density $\Delta
F/V$ between the superconducting phase and the normal phase as a
function of temperature $T$ for two different values of the
external magnetic field $H$. As can be seen, the free energy
density differences are smooth at the critical temperatures (the
first derivatives of the free energy differences are continuous at
$T_c$) which suggests that the phase transitions at the critical
temperatures are of second order. When the magnetic field
vanishes, $H=0$, the critical temperature $T_c=T_{c0}$, and we get
similar results which are consistent with those given in
\cite{roberts}.

\section{ conclusion and discussion}

In this paper, we have studied the effects of an external magnetic
field for a holographic superconductor dual to an EYM theory. A
critical magnetic field exists for this model. Below this critical
magnetic field, a condensate develops and triggers
superconductivity. The critical magnetic field corresponds to the
maximal magnetic charge the black hole can carry such that the
black hole is censored by a horizon. This is consistent with the
concept of the hairy black hole that the spontaneous symmetry
breaking of the gauge invariance which results in
superconductivity occurs slightly outside the horizon
\cite{gubser2}.

The existence of a critical magnetic field at a given critical
temperature for the holographic superconductor considered in this
work is expected for real-world superconductors. The critical
temperature drops as the external magnetic field increases.
Similar conclusion is obtained by considering a rotating
holographic superconductor with a dual gravity background given by
the Kerr-Newman-AdS solution \cite{sonner}. The critical
temperature is found to decrease as the angular momentum of the
dual black hole increases. It would be interesting to consider the
effects of adding both magnetic field and angular momentum to the
dual black hole.

The existence of a superconductivity phase in spite of the presence
of an external magnetic field is a signal of the Meissner effect.
However, our simple model can not distinguish between type
\Rmnum{1} and type \Rmnum{2} superconductors. The ansatz (6)
considered in this paper has a symmetric form for the $SU(2)$
gauge fields $A_1^1=A_2^2=w$. Interesting results may be obtained
by considering a less symmetric ansatz such as
$A_1^1=w,A_2^2=0$, the one studied in \cite{gubser4}.

\section{acknowledgement}
We thank professors S.~A.~Hartnoll and C.~P.~Herzog for great help.
We also thank G.~Chen, P.~Kerner and J.~Sonner for helpful comments.
This work is supported by the National Natural Science Foundation of
China under Grants No. 10535010, 10675090, 10775068, and 10735010.

\end{document}